\documentclass[aps,prd,secnumarabic,nobibnotes,twocolumn,superscriptaddress]{revtex4-1}
\usepackage{amsfonts}
\usepackage{mathrsfs}
\usepackage{amsmath}
\usepackage{color}
\usepackage{natbib}
\usepackage{graphicx}
\usepackage{bm}
\usepackage{amssymb}
\usepackage{xspace}
\usepackage{epstopdf}
\usepackage{dcolumn}
\usepackage{multirow}
\usepackage[colorlinks=true, letterpaper=true, pdfstartview=FitV, linkcolor=blue, citecolor=blue, urlcolor=blue]{hyperref}
\usepackage{wrapfig}

\makeatletter

\newcommand{\Rmnum}[1]{\expandafter\@slowromancap\romannumeral #1@}
\makeatother

\begin{document}
\title{Perovskite-type YRh$_{3}$B with multiple types of nodal point and nodal line states}

\author{Feng Zhou}
\address{School of Physical Science and Technology, Southwest University, Chongqing 400715, China}

\author{Chaoxi Cui}
\address{Key Lab of Advanced Optoelectronic Quantum Architecture and Measurement (MOE), Beijing Key Lab of Nanophotonics $\&$ Ultrafine Optoelectronic Systems, and School of Physics, Beijing Institute of Technology, Beijing 100081, China}

\author{Jianhua Wang}
\address{School of Physical Science and Technology, Southwest University, Chongqing 400715, China}
\author{Minquan Kuang}
\address{School of Physical Science and Technology, Southwest University, Chongqing 400715, China}
\author{Tie Yang}
\address{School of Physical Science and Technology, Southwest University, Chongqing 400715, China}

\author{Zhi-Ming Yu}
\address{Key Lab of Advanced Optoelectronic Quantum Architecture and Measurement (MOE), Beijing Key Lab of Nanophotonics $\&$ Ultrafine Optoelectronic Systems, and School of Physics, Beijing Institute of Technology, Beijing 100081, China}
\author{Xiaotian Wang}\email{xiaotianwang@swu.edu.cn}
\address{School of Physical Science and Technology, Southwest University, Chongqing 400715, China}
\author{Gang Zhang}\email{zhangg@ihpc.a-star.edu.sg}
\address{Institute of High Performance Computing, Agency for Science, Technology and Research (A*STAR), 138632, Singapore}
\author{Zhenxiang Cheng}\email{cheng@uow.edu.au}
\address{Institute for Superconducting and Electronic Materials (ISEM), University of Wollongong, Wollongong 2500, Australia}

\date{December 27, 2020}

\begin{abstract}
Experimentally synthesized perovskite-type YRh$_{3}$B with a $Pm\bar{3}m$ type structure was proposed as a novel topological material (TM) via first-principles calculations and the low-energy $k\cdot p$ effective Hamiltonian, which has a quadratic contact triple point (QCTP) at point $\Gamma$ and six pairs of open nodal lines (NLs) of the hybrid type. Clear surface states observed in the surface spectrum confirmed the topological states. When spin-orbit coupling was considered, the QCTP at $\Gamma$ transferred to the quadratic-type Dirac nodal point (NP). Under 1$\%$ tetragonal strained lattice constants, YRh$_{3}$B hosted richer topological states, including a quadratic-type two-fold degenerate NP, six pairs of open NLs of the hybrid type, and two closed NLs of type I and hybrid type. Moreover, it was proved that the NLs of YRh$_{3}$B at its strained lattice constants contain all types of band-crossing points (BCPs) (i.e., type I, type II, and critical type). Such rich types of NP and NL states in one compound make it potentially applicable for multifunctional electronic devices as well as an appropriate platform to study entanglement among topological states.
\end{abstract}
\pacs{Valid PACS appear here}
\maketitle

\section{\label{sec:level1}Introduction}
Topological semimetals/metals~\cite{add1,add2,add3,add4,add5}, have attracted tremendous research attention due to the nontrivial band-crossings in their low-energy states, which give rise to novel fermionic excitations. Basically, topological semimetals/metals include three types: nodal-point~\cite{add6,add7,add8,add9,add10}, nodal-line~\cite{add11,add12,add13,add14,add15}, and nodal-surface topological semimetals/metals~\cite{add16,add17,add18}. Regarding nodal-surface topological materials (TMs), their classification, electronic structures, and material realization were studied preliminarily by Liang \textit{et al}.~\cite{add17} and Wu \textit{et al}.~\cite{add18}. However, related research is still in its infancy, and much in-depth physical research needs to be conducted.

Nodal-point materials can be classified in at least the following methods: (i) The degeneracy of the point. For example, recently, significant efforts have been made to propose nodal point (NP) materials with 2-, 3-, 4-, 6-, and 8-fold degenerated band crossings~\cite{add4,add7,add19,add20,add21,add22,add23,add24}. Among them, NP materials with 2-, 3-, and 4-fold degenerated band crossings---i.e., Weyl, triply degenerated, and Dirac NP materials---are of significant importance due to their novel physical properties; (ii) The band dispersion near the points. Besides the most common linear dispersion, in principle, a quadratic and a cubic dispersion may also be found around NPs in the low-energy regions~\cite{add25,add26,add27,add28,add29}. Moreover, the linear dispersion around the NPs can also be classified into three types according to the tilting degree of the fermion cone: type I~\cite{add30,add31,add32}, type II~\cite{add33,add34,add35}, and critical-type~\cite{add14}. In type I (type II) NPs, the two bands, which form the NP, have opposite (equal) velocities; meanwhile, critical-type NPs are formed by one flat band and one dispersive band~\cite{add14}.

For nodal line (NL) materials, the BCPs can form NL states in the momentum space of solids. If all the BCPs in the NL belong to one type, the NL is of the corresponding type~\cite{add36,add37}; meanwhile, if the NL contains one more type of BCP, the NL is hybrid-type~\cite{add38}. Additionally, Yu \textit{et al}.~\cite{add39} found that higher-order NLs can also be found in nonmagnetic systems. More interestingly, NL states in momentum space can form many sub-types. For example, NLs can be classified into two types---closed NLs and open NLs---based on whether the NLs are expanded to the edge of the first Brillouin zone (BZ)~\cite{add40,add41}.

In this work, we report an interesting TM with rich types of NL and NP states under equilibrium and strained lattice constants. First, we show that monoxide perovskite-type compound YRh$_{3}$B is a dynamically stable material with a triply degenerate NP with higher-order quadratic dispersion at $\Gamma$ and six pairs of hybrid open NLs. Second, with spin-orbit coupling (SOC), we demonstrate that the quadratic contact triple point (QCTP) - quadratic contact Dirac point (QCDP) transition occurs at point $\Gamma$. Third, under 1$\%$ tetragonal strain, a quadratic-type two-fold NP, six pairs of open NLs, and two closed NLs appeared in the momentum space of YRh$_{3}$B. Then, we reveal that BCPs of type I, type II, and critical type can be observed at strained lattice constants.

\begin{figure}
\includegraphics[width=7cm]{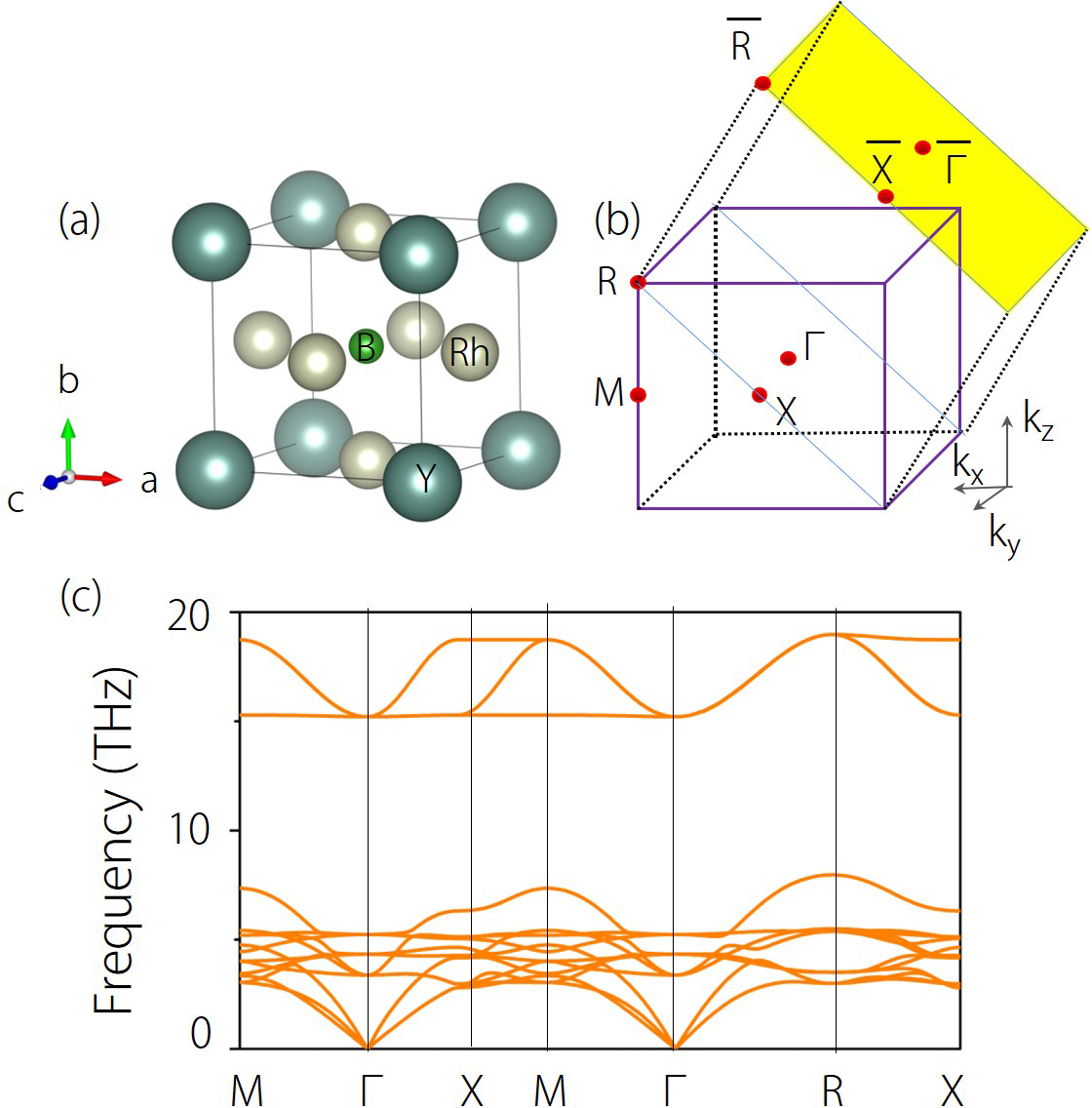}
\caption{(a) The crystal structure of YRh$_{3}$B. (b) The three-dimensional (3D) bulk Brillouin zone (BZ) and the surface BZ of the [101] surface (shown in yellow). (c) The calculated phonon dispersion of YRh$_{3}$B at its equilibrium lattice constant along M-$\Gamma$-X-M-$\Gamma$-R-X paths.
\label{fig1}}
\end{figure}

\section{\label{sec:level1} MATERIALS AND COMPUTATIONAL METHODS}

In this work, a first-principles method, as implemented in the Vienna ab initio Simulation Package~\cite{add42}, was selected to calculate the electronic structure and the topological properties of YRh$_{3}$B at equilibrium and strained lattice constants. We adopted the generalized gradient approximation (GGA)~\cite{add43} of the Perdew--Burke--Ernzerhof (PBE)~\cite{add44} function as the exchange-correlation potential. Moreover, we set the cutoff energy as 600 eV and sampled the BZ via a Monkhorst--Pack $\mathit{k}$-mesh with a size of 10 $\times$ 10 $\times$ 10. We set the energy/force convergence criteria as 10$^{-6}$ eV/ -0.0005 eV/\AA. To achieve the phonon dispersions of YRh$_{3}$B systems, a 2 $\times$ 2 $\times$ 2 supercell was selected and the phonon spectra were calculated via the force-constants method as implemented in Phonopy code~\cite{add45}. The surface states are calculated by using Wannier functions~\cite{add46}. The intensity is calculated by Green's function method for the [101] surface~\cite{add47,add48}.

YRh$_{3}$B sample was prepared via the arc-melting method in a titanium-gettered argon atmosphere before~\cite{add49}. In this work, the lattice constants of YRh$_{3}$B were totally relaxed via first-principles calculation, and the obtained values were a = b = c = 4.1996 \AA, which are in good agreement with the experimental values~\cite{add49}, namely a = b = c = 4.22 \AA. The optimized crystal structure of YRh$_{3}$B at its equilibrium lattice constants is shown in Fig.~\ref{fig1}(a). This primitive cell has five atoms, namely one B atom, one Y atom, and three Rh atoms, which are located at the 1b (0.5, 0.5, 0.5), 1a (0.0, 0.0, 0.0), and 3c (0.5, 0.0, 0.5) Wyckoff sites, respectively.

\begin{figure}
\includegraphics[width=8.8cm]{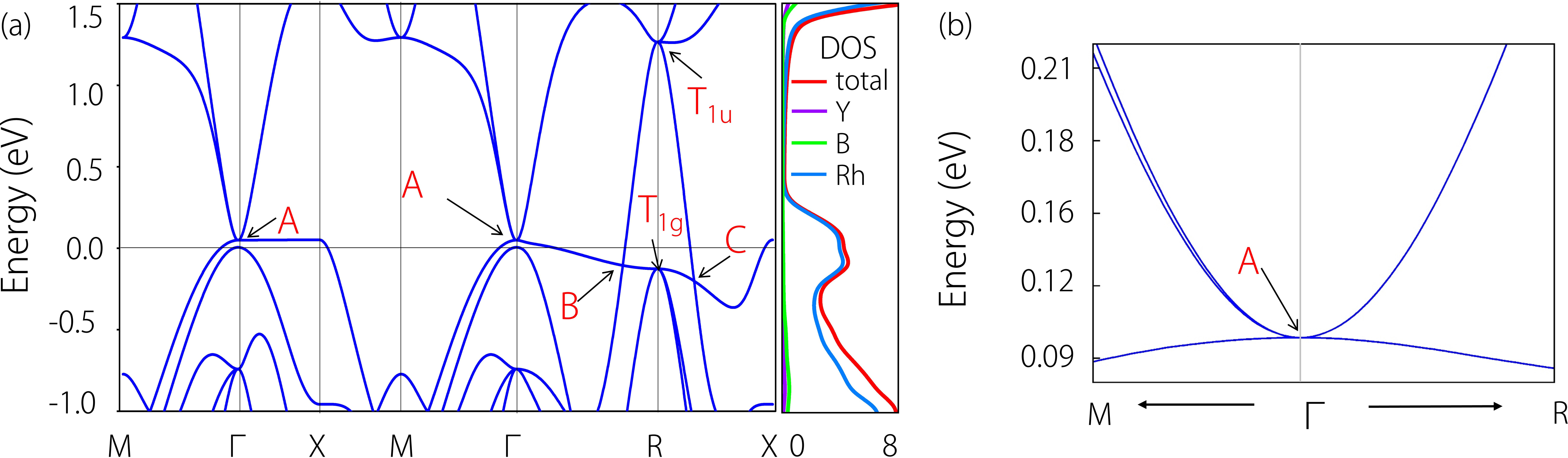}
\caption{(a) The band structure and the (total and partial) DOSs of YRh$_{3}$B at its equilibrium lattice constant along M-$\Gamma$-X-M-$\Gamma$-R-X paths. (b) The enlarged band structure of YRh$_{3}$B along M-$\Gamma$-R paths; point A is a triply degenerated NP with a quadratic manner.
\label{fig2}}
\end{figure}

To evaluate the mechanical stability of YRh$_{3}$B, the elastic constants were obtained. For cubic-system YRh$_{3}$B, there are three independent elastic constants, namely, $C_{11},$  $C_{12}$ and $ C_{44}.$ The values of $C_{11},$  $C_{12}$ and $ C_{44}$, are 333.59 GPa, 102.78 GPa and 68.16 GPa, respectively. The elastic stability criteria are as follows:

\begin{equation}
    C_{11}-C_{12}>0,
\end{equation}
\begin{equation}
    C_{11}+2C_{12}>0,
\end{equation}
\begin{equation}
    C_{44}>0.
\end{equation}
The elastic constants of YRh$_{3}$B satisfy all of the above criteria~\cite{add50}, and therefore the material is mechanically stable.

The Bulk modulus ($\mathit{B}$) is calculated based on the following formula:

\begin{equation}
    B_{cal.}=\frac{\left(C_{11}+2C_{12}\right)}{3}.
\end{equation}
The calculated value of $\mathit{B_{cal.}}$ is 181.72 GPa, which is very close to the experimental value ($\mathit{B_{exp.}}$ = 177.4 GPa)~\cite{add49}.

\begin{figure}
\centering
\includegraphics[width=8cm]{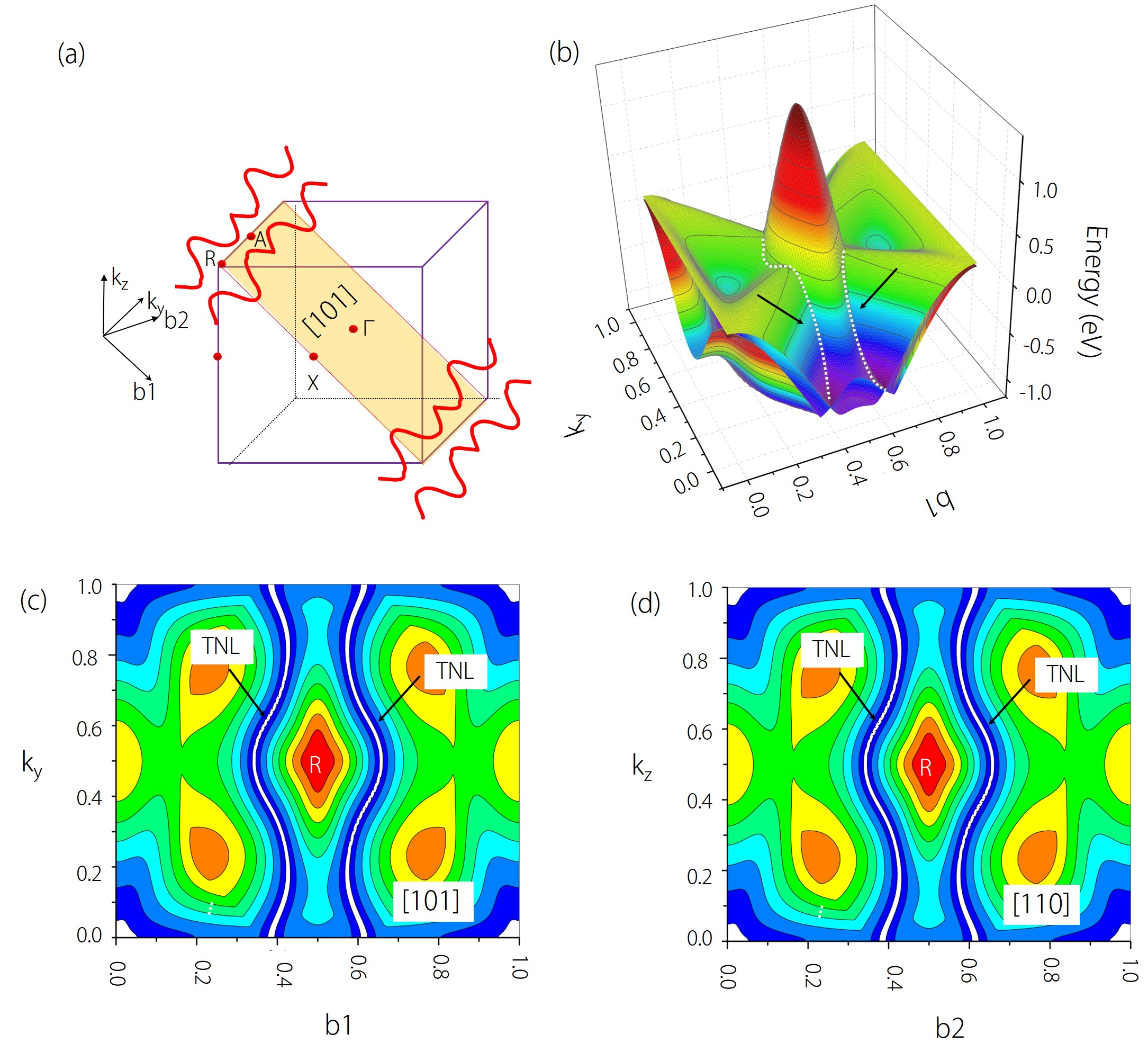}
\caption{(a) A schematic diagram of one pair of open nodal lines (NLs) (red lines) in the [101] plane. (b) The 3D band dispersion of the [101] plane around point R. The white dotted lines represent the NL states in the [101] plane. (c) and (d) The shapes of the topological NLs (TNLs) (white dotted lines) of the [101] and [110] planes.
\label{fig3}}
\end{figure}

\begin{figure}
\includegraphics[width=8cm]{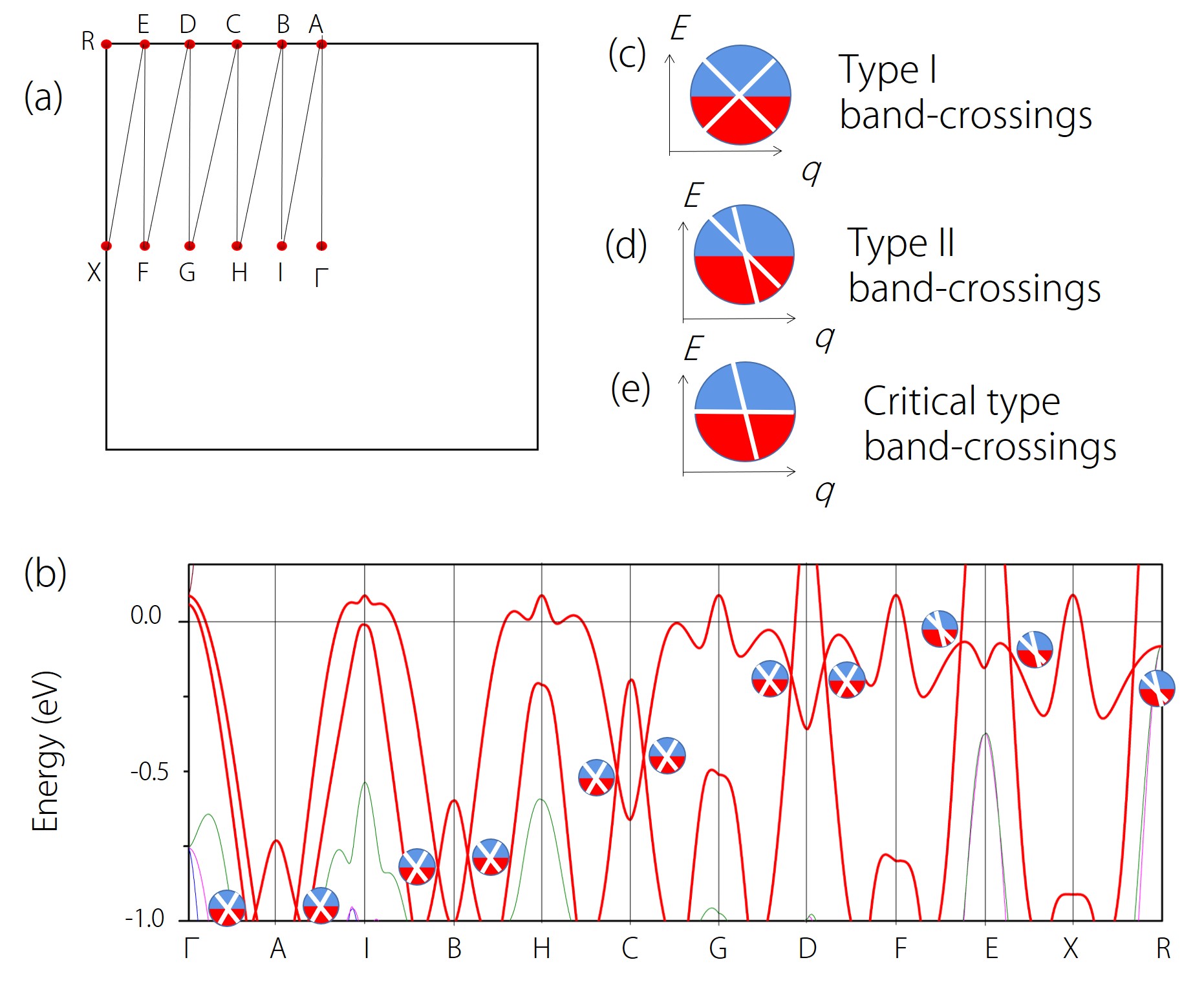}
\caption{(a) Selected $k$-paths in the [101] plane; points B, C, D, E (F, G, H, I) are equally spaced between R (X) and A ($\Gamma$). (b) Calculated band structures along the $\Gamma$-A, A-I, I-B, B-H, H-C, C-G, G-D, D-F, F-E, E-X, and X-R paths, respectively. (c)--(e) An illustration of NPs with type I, type II, and critical-type band crossings.
\label{fig4}}
\end{figure}

The phonon spectrum is widely used to study the stability of the system~\cite{add51,add52} and the occurrence of imaginary eigenfrequencies generally indicates a structural instability~\cite{add53,add54}. Hence, we will use this property to study the the dynamically stability of YRh$_3$B. The calculated phonon dispersion of YRh$_3$B at its equilibrium lattice constants is exhibited in Fig.~\ref{fig1}(c). No imaginary frequency mode in phonon dispersion indicates that this compound is dynamically stable.

\section{\label{sec:level1}ELECTRONIC AND TOPOLOGICAL SIGNATURES AT EQUILIBRIUM LATTICE CONSTANTS}

Fig.~\ref{fig2}(a) shows the calculated band structure and the (total and partial) DOSs of YRh$_{3}$B at its equilibrium lattice constant without SOC. Based on the band structure, it can be seen that this compound is metallic and has several remarkable crossings around the Fermi level, which are labeled as A, B and C. We first consider point A. The enlarged band structure around A is presented in Fig.~\ref{fig2}(b), which shows that this point is triply degenerated~\cite{add55,add56} and that the three bands completely split along the $\Gamma$-M path and become a doubly degenerate NL and a single band along the $\Gamma$-R path. The triple point at A ($\Gamma$) is a quadratic contact triple point (QCTP) as its band representation is $T_{1g}$\ of the $O_h$ point group. Specifically, the little group of point $\Gamma$ is the $O_h$ point group, and its generating elements can be chosen as $C_{4z}$, $M_{110}$, $C_{3,111}$, ${\cal P}$ and time-reversal ${\cal T}$. With these symmetries constrained, the low-energy $k\cdot p$ effective Hamiltonian of point A can be calculated as follows:
\begin{equation}
\begin{aligned}
   {\cal H}_{\text{QCTP}}=c_{1}+&c_{2}k^{2}+c_{3}\left(\begin{array}{ccc}
k_{z}^{2} & 0 & 0\\
0 & k_{y}^{2} & 0\\
0 & 0 & k_{x}^{2}
\end{array}\right)\\
&+c_{4}\left(\begin{array}{ccc}
0 & k_{y}k_{z} & -k_{x}k_{z}\\
k_{y}k_{z} & 0 & k_{x}k_{y}\\
-k_{x}k_{z} & k_{x}k_{y} & 0
\end{array}\right),\label{eq:Gammaham}
\end{aligned}
\end{equation}
where $c$'s are real parameters. One can check that the effective Hamiltonian (Eq. 5) recovers the electronic band structure in Fig.~\ref{fig2}. Moreover, from the effective Hamiltonian, one can find that point A has quadratic dispersion along any direction in the BZ; hence, this point is termed a QCTP. In particular, along the three momentum axes (e.g., along the $k_{x}$ direction), the energy dispersion of QCTP is as follows:
\begin{equation}
\varepsilon_{1,2}=c_{1}+c_{2}k_{x}^{2},\ \ \varepsilon_{3}=c_{1}+(c_{2}+c_{3})k_{x}^{2}.
\end{equation}
Then, we find $c_{2}\simeq-c_{3}$ in YRh$_{3}$B, leading to the appearance of an almost flat band along the $k_{x,y,z}$ direction (see the example of [$\Gamma$-X path] in Fig.~\ref{fig2}).

\begin{figure}
\includegraphics[width=8cm]{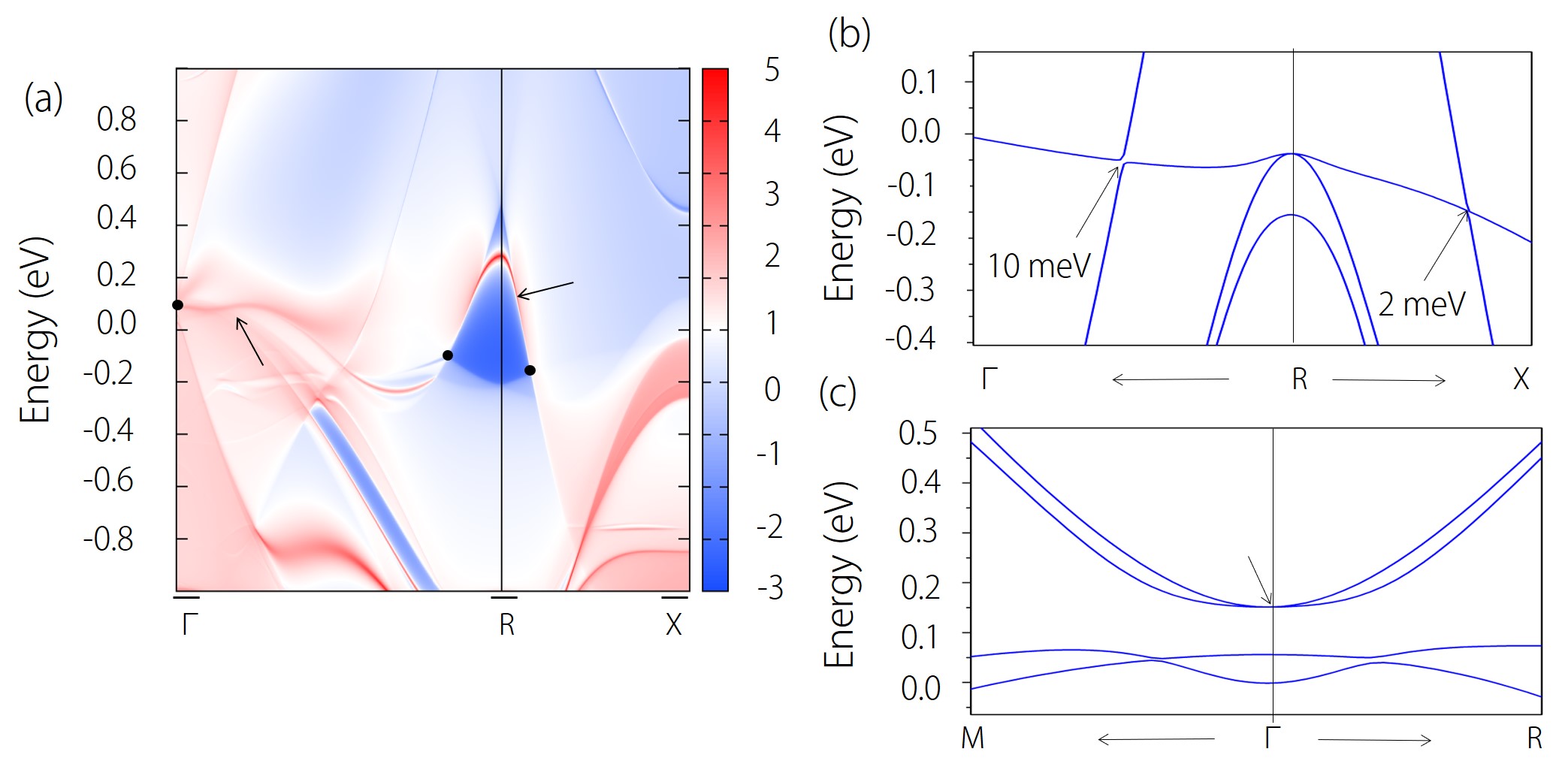}
\caption{(a) The [101] surface states of YRh$_{3}$B along the surface Brillouin zone $\bar{\Gamma}-\bar{\rm{R}}-\bar{\rm{X}}$; the positions of points A, B, C (as shown in Fig.~\ref{fig2}) are marked by black dots and the surface sates arising from these points are marked by arrows. The colorscale represents intensity on a log scale. (b) and (c) The calculated band structures of YRh$_{3}$B along the $\Gamma$-R-X and M-$\Gamma$-R directions, respectively, with the spin-orbit coupling (SOC) effect.
\label{fig5}}
\end{figure}

We then consider points B and C, which are accidental band crossings appearing at high-symmetry lines $\Gamma$-R and R-X, respectively. Since the YRh$_{3}$B material has
${\cal P}$ and ${\cal T}$ symmetries, points B and C should not be isolated according to the arguments by Weng \textit{et al}~\cite{add36,add57}. As both the $\Gamma$-R and R-X paths occur at the [101] mirror plane, we studied the band structure of YRh$_{3}$B in the [101] plane in detail by plotting a 3D band structure around R (see Fig.~\ref{fig3}), which directly shows that points B and C are not isolated but reside on an open NL crossing the BZ around the R-A path, as shown in Fig.~\ref{fig3}(a). Due to the ${\cal P}$ symmetry of the YRh$_{3}$B material, there exists one pair of NLs rather than one open NL around the R-A path (see Fig.~\ref{fig3}). For clarity, a more detailed calculation was performed for the YRh$_{3}$B material at its equilibrium lattice constant. As shown in Fig.~\ref{fig4}(a), we selected 11 other $\mathit{k}$-paths throughout the NL states to perform the calculations of band structures. The selected $\mathit{k}$-paths are the $\Gamma$-A, A-I, I-B, B-H, H-C, C-G, G-D, D-F, F-E, E-X, and X-R paths, respectively. The band structures of the above-mentioned $\mathit{k}$-paths are given in Fig.~\ref{fig4}(b). From the figure, it can be seen that the linear BCPs appear along all of the 11 selected $\mathit{k}$-paths. Additionally, based on the tilting degree of the band dispersion around the points in Fig.~\ref{fig4}(c)--(e), the point on open NL transfers from type I to type II from path $\Gamma$-A to path X-R. Moreover, since YRh$_{3}$B has an $O_h$ point group, there are five more pairs of such open NL states in [110], [011], [$\bar{1}$10], [01$\bar{1}$],[10$\bar{1}$], which are all connected to the [101] mirror by symmetry. As an example, in Fig.~\ref{fig3}(d), one pair of open NL states are exhibited in [110] plane.

Since the low-energy part of the open NL is mainly located around point R, we establish a $k\cdot p$ effective Hamiltonian at point R to capture the low-energy feature of the open NL. From the electronic bands of YRh$_{3}$B (see Fig.~\ref{fig2}), it can be found that the formation of the open nodal ring involves six entangled bands, which are degenerated into one pair of triply degenerate points. As YRh$_{3}$B only has symmorphic symmetries, the little group of point R also is the $O_h$ point group, and the irreducible representations of the two triply degenerate points at point R around the Fermi level are $T_{1g}$ and $T_{1u}$, respectively, as shown in Fig.~\ref{fig2}. Using $T_{1g}$ and $T_{1u}$ representations as a basis, the $k\cdot p$ model (up to second-order) can be expressed as follows:

\begin{equation}
{\cal H}_{R}=\left[\begin{array}{cc}
h_{11} & h_{12}\\
h_{12}^{\dagger} & h_{22}
\end{array}\right],
\end{equation}
where each block is a $3\times3$ submatrix with

\begin{equation}
\begin{aligned}
h_{11}=M_{1}+&E_{1}k^{2}+D_{1}\left(\begin{array}{ccc}
k_{z}^{2} & 0 & 0\\
0 & k_{y}^{2} & 0\\
0 & 0 & k_{x}^{2}
\end{array}\right)\\
&+A_{1}\left(\begin{array}{ccc}
0 & k_{y}k_{z} & -k_{x}k_{z}\\
k_{y}k_{z} & 0 & k_{x}k_{y}\\
-k_{x}k_{z} & k_{x}k_{y} & 0
\end{array}\right),
\end{aligned}
\end{equation}

\begin{equation}
\begin{aligned}
h_{22}=M_{2}+&E_{2}k^{2}+D_{2}\left(\begin{array}{ccc}
k_{x}^{2} & 0 & 0\\
0 & k_{y}^{2} & 0\\
0 & 0 & k_{z}^{2}
\end{array}\right)\\
&+A_{2}\left(\begin{array}{ccc}
0 & k_{y}k_{x} & k_{x}k_{z}\\
k_{y}k_{x} & 0 & k_{z}k_{y}\\
k_{x}k_{z} & k_{z}k_{y} & 0
\end{array}\right),
\end{aligned}
\end{equation}
and

\begin{equation}
h_{12}=iB\left(\begin{array}{ccc}
-k_{y} & k_{x} & 0\\
-k_{z} & 0 & k_{x}\\
0 & -k_{z} & k_{y}
\end{array}\right),
\end{equation} where the wave vector $\boldsymbol{k}$ is measured from point R point $M_{i}$,
$D_{i}$, $E_{i}$, $A_{i}$ with ($i=1,2$) and\textbf{ }$B$ are the real coefficients related to the specific material. Both $h_{11}$
and $h_{22}$ describe a QCTP, and the crossings between them around the Fermi level form the open NL. Note that, due to the three-fold rotation along the $\Gamma$-R direction, point B (see Fig.~\ref{fig2}(a)) actually resides at the intersection point of three symmetry-related open NL.

\begin{figure}
\includegraphics[width=8cm]{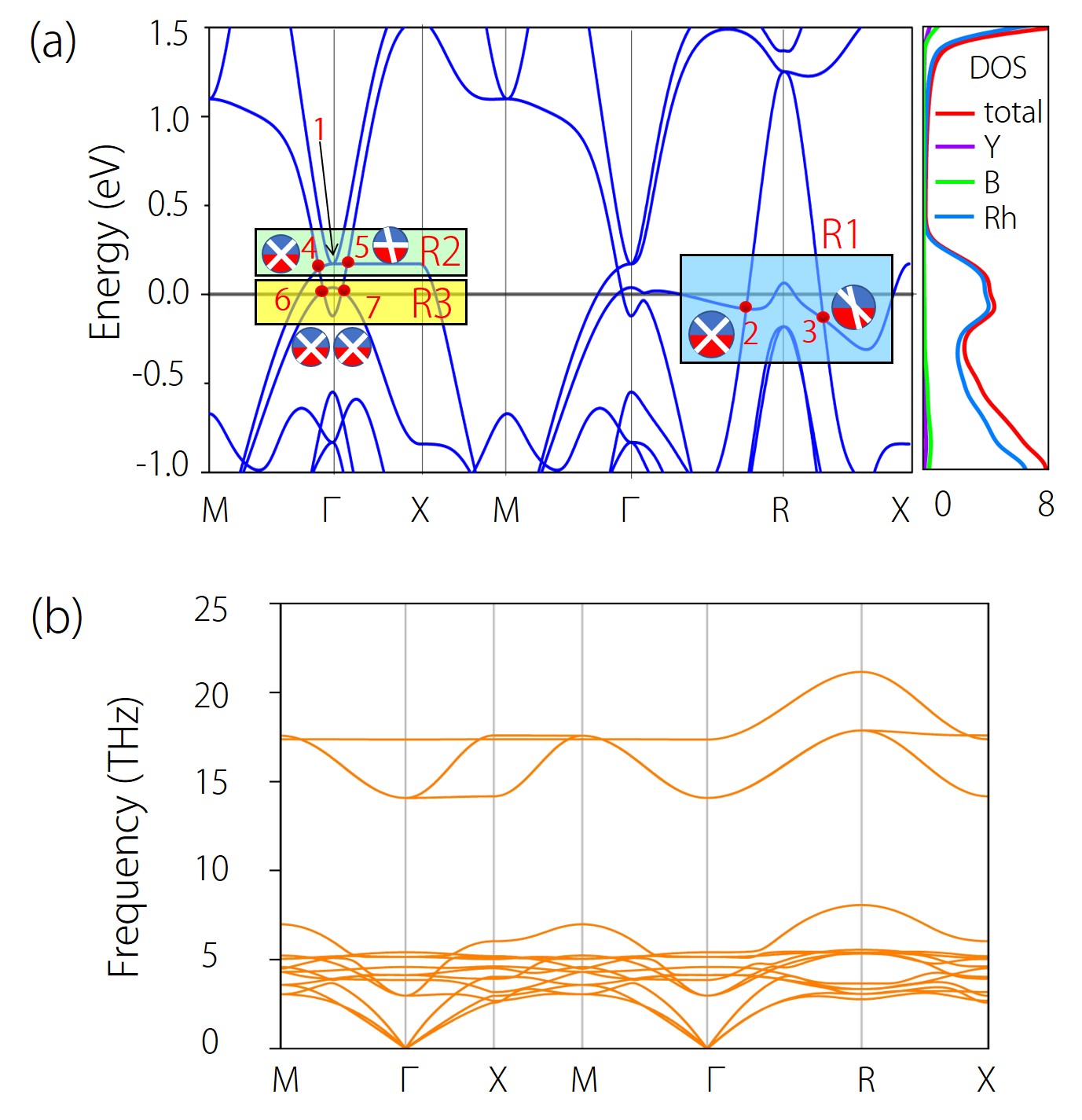}
\caption{(a) The band structure and the (total and partial) DOSs  of YRh$_{3}$B with 1$\%$ tetragonal strain and without SOC; the bands around point 1 are with a quadratic manner at point $\Gamma$. For the other numbered linear BCPs, their types and positions are marked. (b) The phonon dispersion of YRh$_{3}$B with 1$\%$ tetragonal strain along M-$\Gamma$-X-M-$\Gamma$-R-X directions.
\label{fig6}}
\end{figure}

The [101] surface spectrum for the NP and NLs is shown in Fig.~\ref{fig5}(a). In the figure, the positions of the NP A and the BCPs, B and C, are marked by black dots, and the surface states originating from these points are shown by black arrows. To reflect the quality of the wannierization of the DFT band structure, the band structures obtained by Wannier along the $\Gamma$-R-X and M-$\Gamma$-R directions, respectively, are shown in Fig. S1(a) and S1(c). Obviously, the band structures around the band-crossing points (A, B, C) obtained by Wannier are in good agreements with those obtained by DFT (see Fig. S1(b) and S1(d)).

\begin{figure}
\includegraphics[width=8cm]{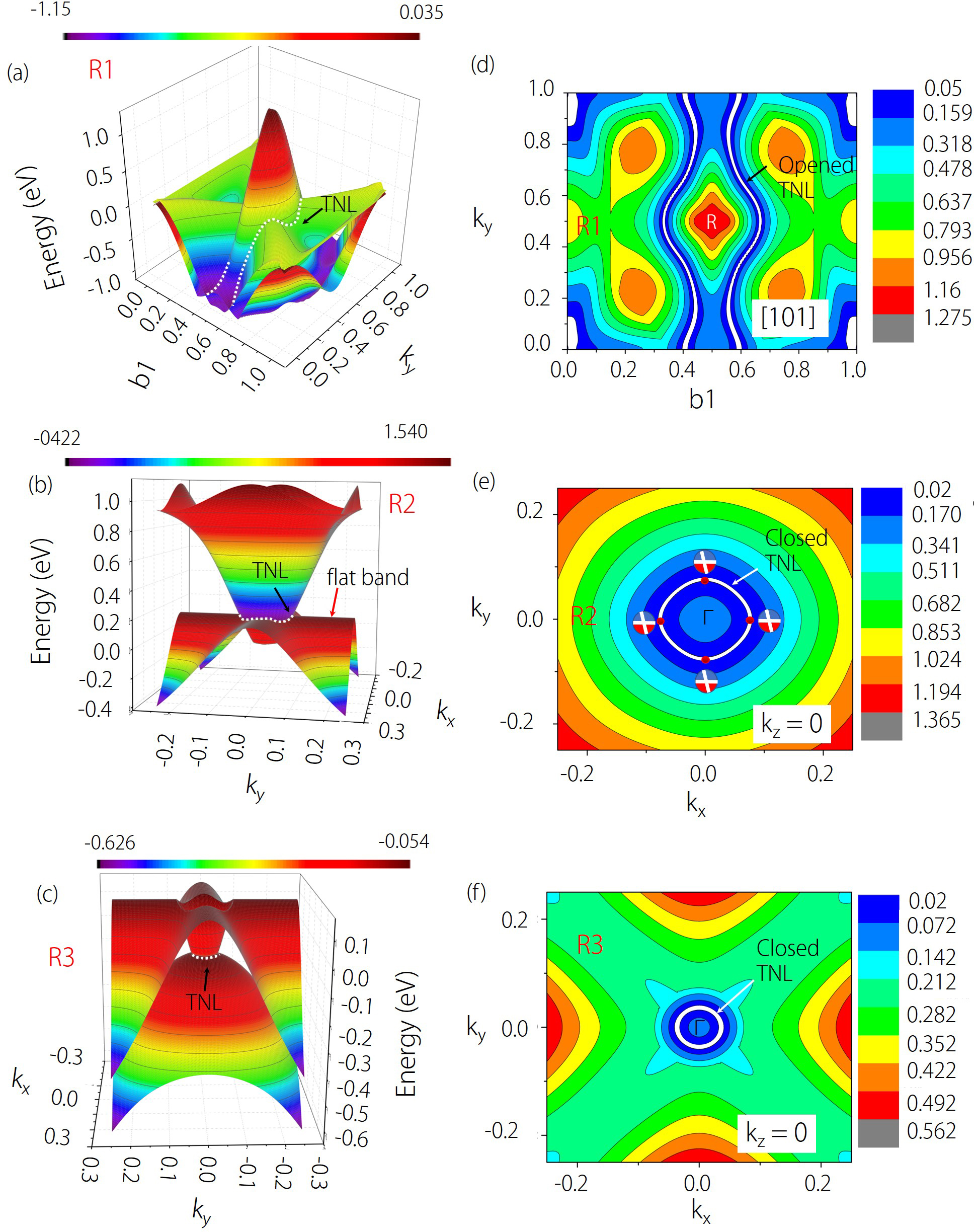}
\caption{(a) The 3D band dispersion of the [101] plane around point R in R1. (b) and (c): The $\Gamma$-centered 3D band dispersion of the $k_z$ = 0 plane in R2 and R3, respectively. (d) The shapes of the open TNLs of the [101] plane. (e) and (f): The shapes of the closed TNLs of the $k_z$ = 0 plane in R2 and R3, respectively. The white dashed and solid lines represent the NL states.
\label{fig7}}
\end{figure}

It should be noted that the topological NL (TNL) states of most predicted NL materials will be opened by the inducing of the SOC effect. Fig.~\ref{fig5}(b) indicates that YRh$_{3}$B is an ideal NL material and its TNL states are very robust to the SOC effect. The SOC-induced gaps for the BCPs B and C along $\Gamma$-R-X paths are up to 10 meV. Furthermore, a more advanced method, modified Becke-Johnson GGA (mBJGGA) functional is also used to calculate the band structures of YRh$_3$B with SOC effect. Note that the mBJGGA functional possesses similar accuracy with more expensive functionals, such as HSE06 and GW, in predicting the band gap, and has been widely used in previous studies ~\cite{add6,add58}. The results are shown in Fig. S2, one can see that the SOC-induced gaps for the BCPs B and C along $\Gamma$-R-X paths are up to 20 meV. It is much smaller than those of well-known NL materials such as ZrSiSe (25-35 meV)~\cite{add59,add60}, Hg$_{3}$As$_{2}$ ($\sim$34 meV)~\cite{add61}, FeSi$_{2}$ (20-30 meV)~\cite{add62}, and BaSn$_{2}$ (50-160 meV)~\cite{add63}. Fig.~\ref{fig5}(c) shows the enlarged band structure of YRh$_{3}$B at point $\Gamma$ when SOC is taken into consideration. From the figure, it can be seen that the QCTP A at point $\Gamma$ is gapped under SOC; however, two higher bands at point $\Gamma$, now contacted with each other at 0.16 eV above the Fermi level. As each band has two-fold degeneracy under SOC, this point is a QCDP. It should be noted that similar triply degenerated NP to Dirac point transition under SOC is also proposed in other TMs with linear-contacted NPs, such as ErAs~\cite{add64}, Li$_{2}$NaN~\cite{add65}, and TiB$_{2}$~\cite{add55}.

\section{\label{sec:level1}  TOPOLOGICAL ELEMENTS AT STRAINED LATTICE CONSTANT}

In this section, we study the electronic structures and the related topological elements of YRh$_{3}$B with 1$\%$ tetragonal strained lattice constants. Now, the space group of YRh$_{3}$B has been transferred from $Pm\bar{3}m$ to $P_4/mmm$ and the lattice constants are a = b = 4.24 \AA ~and c = 4.11 \AA, respectively. The phonon dispersion of YRh$_{3}$B at 1$\%$ tetragonal strained lattice constants is shown in Fig.~\ref{fig6}(b) to demonstrate the material's dynamical stability. The band structure and the (total and partial)
DOSs without SOC along M-$\Gamma$-X-M-$\Gamma$-R-X directions is shown in Fig.~\ref{fig6}(a). From the figure, it can be seen that the physical nature of this compound is still metallic under 1$\%$ tetragonal strain, and seven points can be found around the Fermi level. It should be noted that the two bands touch in a quadratic manner at point $\Gamma$ (see point 1) with two-fold degeneracy. Meanwhile, the bands around BCPs 2--7 are with a linear manner in low-energy regions. We divided these BCPs into three regions, which are labeled as R1, R2, and R3, respectively.

\begin{figure}
\includegraphics[width=8cm]{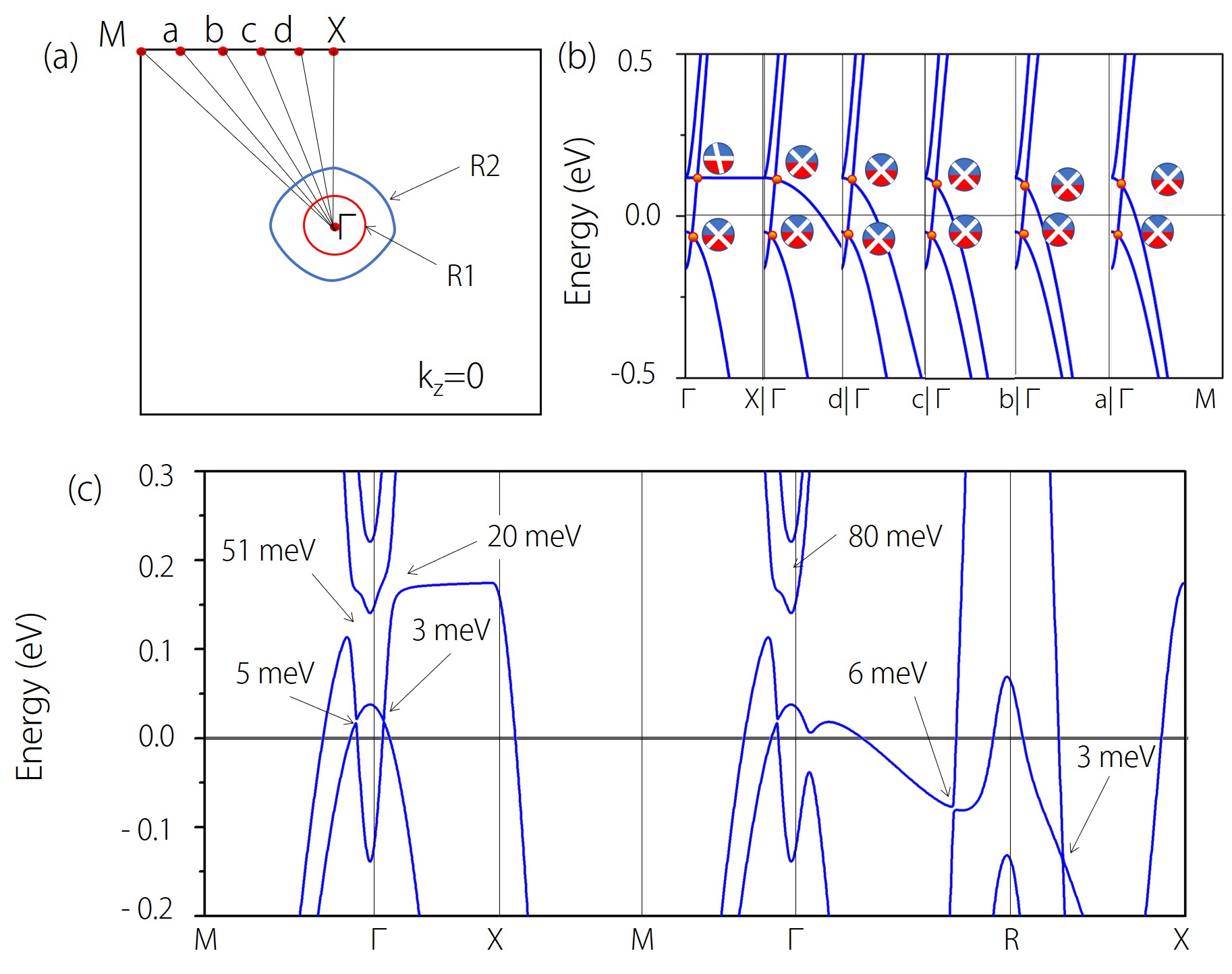}
\caption{(a) Selected $k$-paths in the $k_z$ = 0 plane; points a, b, c, d are equally spaced between M and X. (b) The calculated band structures along the $\Gamma$-M, $\Gamma$-a, $\Gamma$-b, $\Gamma$-c, $\Gamma$-d, and $\Gamma$-X paths, respectively. (c) The calculated band structure of YRh$_{3}$B with 1$\%$ tetragonal strain along the M-$\Gamma$-X-M-$\Gamma$-R-X directions. SOC is considered.
\label{fig8}}
\end{figure}

For BCPs 2 and 3 in R1, they belong to the open NLs in the [101] plane, as shown in Fig.~\ref{fig7}(a) and Fig.~\ref{fig7}(d). Similar to the discussion in Section III, it can be seen that the YRh$_{3}$B compound at 1$\%$ tetragonal strained lattice constants still hosts six pairs of open NL states in the [101], [110], [011], [$\bar{1}$10], [01$\bar{1}$], and [10$\bar{1}$] planes.

Besides the above-mentioned topological elements, there are four other BCPs in R2 and R3---two along the M-$\Gamma$ path and two along the $\Gamma$-X path. These BCPs should belong to some nodal structures since the ${\cal P}$ and ${\cal T}$ symmetries are still maintained in this system. Very interestingly, the BCPs in Fig.~\ref{fig6}(a) are of three parts: point 3 is of type II; points 2, 4, 6, and 7 are of type I; and point 5 is of the critical type~\cite{add14}. Note that, some previous works reported that the existence of both type I and type II BCPs in realistic materials~\cite{add66,add67}.  However, all-round types (type I, type II and critical type) of BCPs in one compound have rarely been reported before. Hence, this system at 1$\%$ tetragonal strained lattice constants can be regarded as a perfect platform to investigate the relationship among BCPs with different tilting degrees.

Based on the band dispersion around point $\Gamma$ in the $k_z$ = 0 plane, as shown in Fig.~\ref{fig7}(b) and Fig.~\ref{fig7}(c), it can be seen that BCPs 4--7 are not isolated and belong to two nodal rings in the $k_z$ = 0 plane. The shape of the two nodal rings in the $k_z$ = 0 plane is given in Fig.~\ref{fig7}(e) and Fig.~\ref{fig7}(f). From these figures, it can be observed that these two nodal rings are closed TNLs. Furthermore, a more detailed band structure calculation was performed for the YRh$_{3}$B compound with 1$\%$ tetragonal strain along selected $k$-paths ($\Gamma$-M, $\Gamma$-a, $\Gamma$-b, $\Gamma$-c, $\Gamma$-d, and $\Gamma$-X; see Fig.~\ref{fig8}(a)). The calculated results, as shown in Fig.~\ref{fig8}(b), indicate that the BCPs were still maintained in the selected $k$-paths. Moreover, from Fig.~\ref{fig8}(b), it can be noted that the $\Gamma$-nodal ring in R1 is type I and the $\Gamma$-nodal ring in R2 is a hybrid type~\cite{add68,add69}. A critical-type BCP to type I BCP transition can be observed from paths $\Gamma$-X to $\Gamma$-M in R2. The critical NP along $\Gamma$-X in R2 is formed by one flat band (see Fig.~\ref{fig7}(b)) and one dispersive band and leads to a new fermionic state. As shown in Fig.~\ref{fig8}(a), the $\Gamma$-nodal ring in R2 has a larger radius than the $\Gamma$-nodal ring in R1.

Hence, it is concluded that YRh$_{3}$B at 1$\%$ tetragonal strained lattice constants is a topological metal with a quadratic type two-fold point at point $\Gamma$, six pairs of open NLs, and two closed nodal rings when SOC is ignored. Then, we come to study the effect of SOC on the band structure and associated topological signatures. The results of band structure with the consideration of SOC are exhibited in Fig.~\ref{fig8}(c). From the figure, it can be seen that the SOC-induced gaps for the BCPs in R1--R3 are very small and the values of these gaps are up to 6 meV. That is, the NL states (including open and closed NLs) in YRh$_{3}$B at 1$\%$ tetragonal strained lattice constants show strong robustness to the SOC effect. However, for the quadratic contacted nodal point at point $\Gamma$, it is broken when the SOC is induced.

\section{\label{sec:level1} REMARKS AND SUMMARY}

In this paper, we propose the realistic material YRh$_{3}$B as a new topological metal. The electronic structures and the topological signatures of YRh$_{3}$B at equilibrium and 1$\%$ tetragonal strained lattice constants are fully investigated based on first-principles calculation. We show that YRh$_{3}$B at equilibrium and 1$\%$ tetragonal strained lattice constants is a topological metallic material with various types of topological signatures. At equilibrium lattice constants, it hosts a triply degenerated NP at point $\Gamma$ with a quadratic manner, and six pairs of open NLs when SOC is ignored. When SOC is considered, a QCTP-to-QCDP transition occurred at point $\Gamma$. At 1$\%$ tetragonal strained lattice constants, there are two bands that touch at point $\Gamma$ near the Fermi level, resulting in a two-fold degenerated NP with a quadratic manner. Besides the two-fold NP, YRh$_{3}$B co-exhibits six open NLs and two closed NLs. Moreover, type I, type II, and critical-type BCPs can be observed in the NLs of YRh$_{3}$B with 1$\%$ tetragonal strain. Our work identifies a novel candidate material to study multiple kinds of NL and NP states in bulk realistic materials. It is hoped that the topological elements proposed in this work can be confirmed in follow-up experimental studies.

\section{\label{sec:level1} ACKNOWLEDGMENTS}

F. Zhou and C. Cui contributed equally to this work. X.T. Wang is grateful for the support from the National Natural Science Foundation of China (No. 51801163) and the Natural Science Foundation of Chongqing (No. cstc2018jcyjA0765).

\end{document}